\documentclass[aps,prl,epsfig,showpacs,superscriptaddress,twocolumn]{revtex4}

\usepackage{graphicx,amsmath,amsfonts,textcomp}

\begin{document}

\title{Towards two-dimensional metallic behavior at LaAlO$_3$/SrTiO$_3$ interfaces}

\author{O. Copie}
\affiliation{Unit\'e Mixte de Physique CNRS/Thales, Campus de l'Ecole Polytechnique, 1 Av. A. Fresnel, 91767 Palaiseau, France and Universit\'e Paris-Sud 11, 91405 Orsay, France}

\author{V. Garcia}
\affiliation{Unit\'e Mixte de Physique CNRS/Thales, Campus de l'Ecole Polytechnique, 1 Av. A. Fresnel, 91767 Palaiseau, France and Universit\'e Paris-Sud 11, 91405 Orsay, France}

\author{C. B\"odefeld}
\affiliation{Attocube System AG, K\"oniginstrasse 11a RGB, D-80539 M\"unchen, Germany}

\author{C. Carr\'et\'ero}
\affiliation{Unit\'e Mixte de Physique CNRS/Thales, Campus de l'Ecole Polytechnique, 1 Av. A. Fresnel, 91767 Palaiseau, France and Universit\'e Paris-Sud 11, 91405 Orsay, France}

\author{M. Bibes}
\affiliation{Unit\'e Mixte de Physique CNRS/Thales, Campus de l'Ecole Polytechnique, 1 Av. A. Fresnel, 91767 Palaiseau, France and Universit\'e Paris-Sud 11, 91405 Orsay, France}

\author{G. Herranz}
\altaffiliation{Now at: Institut de Ci\`encia de Materials de Barcelona, CSIC, Campus de la UAB, 08193 Bellaterra, Cataluyna, Spain}
\affiliation{Unit\'e Mixte de Physique CNRS/Thales, Campus de l'Ecole Polytechnique, 1 Av. A. Fresnel, 91767 Palaiseau, France and Universit\'e Paris-Sud 11, 91405 Orsay, France}

\author{E. Jacquet}
\affiliation{Unit\'e Mixte de Physique CNRS/Thales, Campus de l'Ecole Polytechnique, 1 Av. A. Fresnel, 91767 Palaiseau, France and Universit\'e Paris-Sud 11, 91405 Orsay, France}

\author{J.-L. Maurice}
\affiliation{Unit\'e Mixte de Physique CNRS/Thales, Campus de l'Ecole Polytechnique, 1 Av. A. Fresnel, 91767 Palaiseau, France and Universit\'e Paris-Sud 11, 91405 Orsay, France}

\author{B. Vinter}
\affiliation{Unit\'e Mixte de Physique CNRS/Thales, Campus de l'Ecole Polytechnique, 1 Av. A. Fresnel, 91767 Palaiseau, France and Universit\'e Paris-Sud 11, 91405 Orsay, France}
\affiliation{Physics Department, University of Nice-Sophia Antipolis, Parc Valrose, 06102 Nice Cedex 2, France}

\author{S. Fusil}
\affiliation{Unit\'e Mixte de Physique CNRS/Thales, Campus de l'Ecole Polytechnique, 1 Av. A. Fresnel, 91767 Palaiseau, France and Universit\'e Paris-Sud 11, 91405 Orsay, France}
\affiliation{Universit\'e d'Evry Val d'Essone, Bd. F. Mitterrand, 91025 Evry, France}

\author{K. Bouzehouane}
\affiliation{Unit\'e Mixte de Physique CNRS/Thales, Campus de l'Ecole Polytechnique, 1 Av. A. Fresnel, 91767 Palaiseau, France and Universit\'e Paris-Sud 11, 91405 Orsay, France}

\author{H. Jaffr\`es}
\affiliation{Unit\'e Mixte de Physique CNRS/Thales, Campus de l'Ecole Polytechnique, 1 Av. A. Fresnel, 91767 Palaiseau, France and Universit\'e Paris-Sud 11, 91405 Orsay, France}

\author{A. Barth\'el\'emy}
\email{agnes.barthelemy@thalesgroup.com}
\affiliation{Unit\'e Mixte de Physique CNRS/Thales, Campus de l'Ecole Polytechnique, 1 Av. A. Fresnel, 91767 Palaiseau, France and Universit\'e Paris-Sud 11, 91405 Orsay, France}

\date{\today}

\begin{abstract}

\vspace{0.5cm}

Using a low-temperature conductive-tip atomic force microscope in cross-section geometry we have characterized the local transport properties of the metallic electron gas that forms at the interface between LaAlO$_3$ and SrTiO$_3$. At low temperature, we find that the carriers do not spread away from the interface but are confined within $\sim$10 nm, just like at room temperature. Simulations taking into account both the large temperature \emph{and} electric-field dependence of the permittivity of SrTiO$_3$ predict a confinement over a few nm for sheet carrier densities larger than $\sim 6\times 10^{13}$~cm$^{-2}$. We discuss the experimental and simulations results in terms of a multi-band carrier system. Remarkably, the Fermi wavelength estimated from Hall measurements is $\sim$16 nm, indicating that the electron gas in on the verge of two-dimensionality.

\end{abstract}

\maketitle

The range of functionalities displayed by transition metal oxides spans from superconductivity to magnetism, ferroelectricity and even multiferroic behavior \cite{imada98,eerenstein2006,bibes2007}, and is further extended by the unexpected effects that may appear at their interfaces \cite{okamoto2004}. Examples include the observation of a ferromagnetic insulating state at the interface between two antiferromagnetic insulators \cite{ueda98}, the prediction of a half-metallic electron gas at the interface between two antiferromagnetic manganites \cite{calderon2008} and the suppression of superconductivity in manganites/cuprate superlattices \cite{hoffmann2005}.

Possibly the most studied oxide interface system so far is the high-mobility metallic state that appears in heterostructures combining two band insulators, namely in LaAlO$_3$//SrTiO$_3$(001) \cite{ohtomo2004}. While doping the SrTiO$_3$ (STO) substrate with oxygen vacancies during growth is known to also produce an extended, bulk-like \cite{frederikse67} high-mobility metallic behaviour \cite{herranz2007b,herranz2006b}, a confined metallic electron gas can be formed at the LAO/STO interface (LAO : LaAlO$_3$) under suitable growth and annealing conditions \cite{thiel2006,reyren2007,basletic2008}. This metallic state has recently been shown to transform into a superconductor below critical temperatures on the order of 200~mK \cite{reyren2007}. Importantly, a clear two-dimensional behavior was found in this superconductive regime. In contrast, while a "quasi-two-dimensional" behavior has been indirectly inferred in the metallic regime \cite{thiel2006}, a comparison between the gas thickness and the Fermi wavelength (determining whether quantum confinement along one direction can emerge or not) has not hitherto been carried out.

In this Letter we report conductive-tip atomic force microscopy (CTAFM) experiments \cite{houze96,planes2001,schneegans2001} at 300K and 8K in cross-section LAO/STO samples and simulations of the charge carrier distribution for different sheet carrier densities. Resistance (R) mappings collected with a lateral resolution of a few nm reveal that a confined conductive metallic gas is present at the interface over a thickness of a few nm both at room \emph{and} low temperature. We show theoretically that when the large electric-field dependence of the permittivity $\chi$ is taken into account, the gas actually remains confined even at low temperature, in agreement with experimental results but in contrast with previous reports \cite{siemons2007}. Finally, we compare the electron gas thickness with the Fermi wavelength and suggest that the interfacial electron gas may indeed be quantized in the growth direction.

The sample studied in this paper is a 2 nm LAO film grown at 750$^\circ$C in an oxygen pressure of 10$^{-6}$ mbar onto a TiO$_2$-terminated STO(001) substrate. After growth, the sample was cooled down to room temperature in 300 mbar of oxygen to avoid doping the whole STO substrate with oxygen vacancies \cite{herranz2007b}. Finally a 200-nm thick amorphous LAO layer was deposited at room temperature onto the epitaxial 2 nm LAO film. To connect the interfacial gas a set of 500 $\mu$m $\times$ 500 $\mu$m squares were etched through the LAO and Al/Au contact pads were lifted off. Standard Hall effect measurements indicated the low temperature sheet carrier density and mobility to be n$_{2D}$=$3\times 10^{13}$~cm$^{-2}$ and $\mu$=650 cm$^2$/Vs, respectively \cite{basletic2008}. A part of the sample was used to prepare a cross-section specimen cutting across some of the Al/Au contacts (see Ref. \onlinecite{basletic2008} for preparation details). To acquire resistance maps during the cross-section CTAFM experiments one of these contacts was connected by silver paste and used to measure the tip-sample resistance.

\begin{figure}[h!]
\centering
\includegraphics[width=\columnwidth]{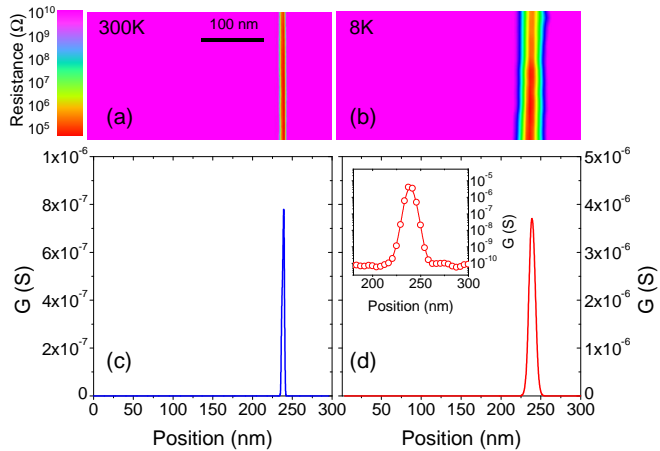}
\caption{(Color online) CTAFM images in cross section geometry collected at 300K (a) and 8K (b). Corresponding conductance profiles (c,d). The inset in (d) shows the same data in logarithmic scale.}
\label{ctafm}
\end{figure}

The room-temperature CTAFM measurements were performed in a Veeco Nanoscope IV setup as described in Ref. \onlinecite{basletic2008}. The low-temperature experiments were carried out in the application lab of Attocube Systems AG using an attoAFM I setup in an exchange gas insert in a liquid He cryostat. In both cases the AFM systems were equipped with a high-performance current measurement module developed by Houz\'e \emph{et al} \cite{houze96}. Nominally identical CrPt-coated tip were used for both types of experiments. The tip-sample current was collected while scanning the sample at a tip-sample bias voltage of -1V (tip-grounded). The topography and resistance were acquired at the same time, which allowed for a precise location of the LAO/STO interface.

The results are summarized in Fig. \ref{ctafm}. At room temperature, the bulk of the STO substrate and the LAO film are fully insulating (R$_{tip-sample} >$10$^{10}$ $\Omega$) and a very thin highly conductive region is present at the LAO/STO interface. At 8K the conductive region has a slightly larger thickness but is still confined close to the interface. A more quantitative estimate of the metallic gas thickness can be inferred from the conductance (G) profiles shown in Fig. \ref{ctafm}c and d. The full width at half-maximum of the G \emph{vs} tip position peak is $\sim$4 nm at 300 K and $\sim$12 nm at 8K. As visible from the inset of Fig. \ref{ctafm}d, the conductance increases by about five orders of magnitude just at the interface.

The very low electron gas thickness measured at 8K is surprising in view of the giant permittivity ($\chi >$ 20000) of STO at low temperature \cite{sakudo71} that has been predicted to spread the carriers away from the interface \cite{siemons2007}. To get more insight into this apparent inconsistency, we have carried out simulations of the carrier distribution in STO taking into account the high non-linearity of the permittivity constant $\chi$ for very large electric fields. We assume a Thomas-Fermi distribution of the carriers, which is appropriate for the cases that follow. Consequently, the carrier density $n(z)$ is linked to the electrical potential $eV(z)=E_{F}-E_{C}(z)$ by

\begin{equation}
n(z)=(2m^{*}/\hbar^{2})^{3/2} [E_{F}-E_{C}(z)]^{3/2}/3\pi^{2}
\end{equation}

\noindent taking into account the spin degeneracy. $E_F$ is the Fermi energy and $E_{C}(z)$ the conduction band profile. $m^{*}$ is the effective mass of the Ti \textit{t$_{2g}$} bands of STO \cite{mattheis72} that is related to both longitudinal ($m_{\parallel}$) and transverse ($m_{\perp}$) electronic masses and to the orbital degeneracy $\nu$ through~\cite{mattheis72}:

\begin{equation}
m^{*}=\nu^{2/3}(m_{\parallel}m_{\perp}^{2})^{1/3}\simeq 4
\end{equation}

We obtain the variation of the band profile $E_{C}$ by solving Poisson's equation $Div[\mathcal{D}]=e \times n(z)$ self-consistently in the continuous limit with $\mathcal{D}=\epsilon_{0}\mathbb{E}+\mathcal{P}$ where $\mathbb{E}=-\frac{1}{e}\frac{\partial E_{C}}{\partial z}$ is the local electric field and $\mathcal{P}=\epsilon_{0} \chi(\mathbb{E}) \mathbb{E}$ the local electrical polarization in STO. Unlike Siemons \emph{et al} \cite{siemons2007}, we assume a field-dependent permittivity constant of the following form:

\begin{equation}
\chi=\frac{\chi_{0}}{1+\mathbb{E}/\mathbb{E}_{C}}
\label{chi-def}
\end{equation}

\noindent where $\mathbb{E}_{C}=\frac{\mathcal{P}_{Sat}}{\epsilon_{0}\chi_{0}}$ accounts for a finite polarization $\mathcal{P}_{Sat}$ in STO~\cite{mattheis72,barrett64}. However, it is important to note that this particular variation form for $\chi$ must not be confused with the one given in the literature~\cite{neville72,christen94,ang2000} that generally expresses the field-dependence of the dynamical permittivity

\begin{equation}
\tilde{\chi}=\frac{1}{\epsilon_0}\frac{\partial P}{\partial \mathbb{E}}
\end{equation}

\noindent in the regime of small electric fields. Quite astonishingly, experimental data give an identical electric field-dependence for $\tilde{\chi}$ according to

\begin{equation}
\tilde{\chi}=\frac{\chi_{0}}{1+\mathbb{E}/\mathbb{E}_{C}^{*}}
\end{equation}

\noindent Using this expression and data from Ang \emph{et al.} at 10~K~\cite{ang2000}, Christen \textit{et al.}~\cite{christen94} and Neville \textit{et al.}~\cite{neville72}, it is possible to extract a value of $\mathbb{E}_{C}^{*}$ of the order of 110 $kV.m^{-1}$ and $\chi_0$=24000. We feel that the particular form of Eq. (\ref{chi-def}) expresses a more physical dependence on $\mathbb{E}$ for $\chi$ at high electric field where $\mathbb{E}_{C}$ and $\mathbb{E}_{C}^{*}$ are linked by

\begin{equation}
\mathbb{E}_{C}=\mathbb{E}_{C}^{*}\ln\{1+\frac{<\mathbb{E}>}{\mathbb{E}_{C}^{*}}\}
\end{equation}

\noindent $<\mathbb{E}>$ being a spatial average of the electric-field amplitude near the LAO/STO interface. From the value of the saturated polarization $\mathcal{P}_{Sat}$ in STO of 0.1~$C.m^{-2}$~\cite{neville72}, we can extract a critical electric field $\mathbb{E}_{C}$ of 470~$kV.$m$^{-1}$, a value that we consider henceforth.

\begin{figure}[h!]
\centering
\includegraphics[width=0.9\columnwidth]{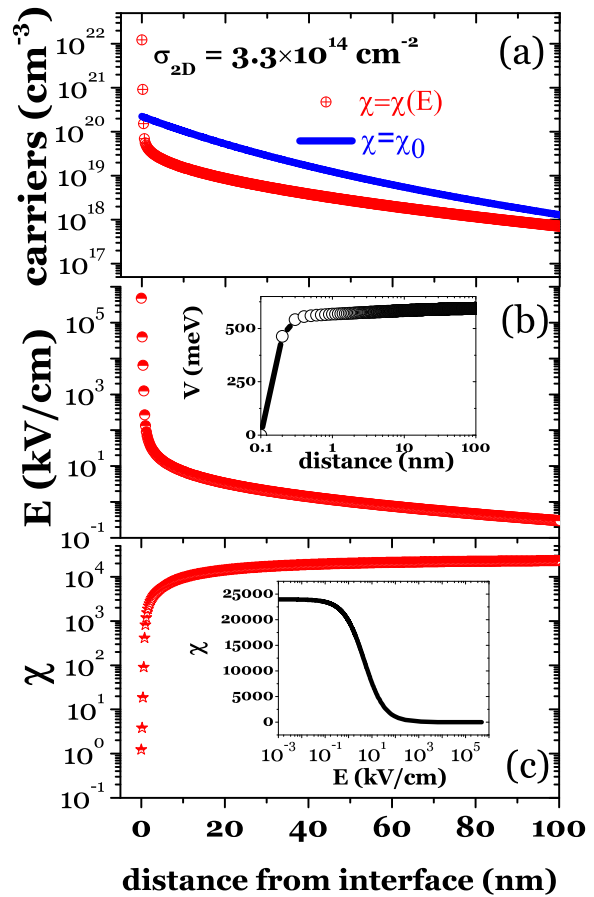}
\caption{(Color online) (a) Carrier density profile at 10 K for n$_{2D}$=$3.3\times 10^{14}$~cm$^{-2}$. Calculations are performed either with $\chi$ constant (as approximated by Siemons \emph{et al}. \cite{siemons2007}) or with a more realistic variation of  $\chi$ with the electric field, see the inset in (c). (b) Dependence of the local electric field with the distance from the LAO/STO interface. The inset shows the corresponding potential profile. (c) Dependence of the dielectric constant with the distance. The electric-field dependence of $\chi$ is shown in the inset.}
\label{profiles}
\end{figure}

Fig. \ref{profiles}a displays the carrier density profiles in STO corresponding to a sheet carrier density of $n_{2D}=3.3\times 10^{14}$~cm$^{-2}$ corresponding to half of an electronic charge per unit cell necessary to avoid the polar catastrophe~\cite{thiel2006,huijben2006}. The latter profile has been calculated at a discretization step of 0.2 nm corresponding to the distance between two atomic planes and in the limit of a vanishing electric field in STO far away from the interface. We either took into account the field-dependence of $\chi$ of Eq.~(\ref{chi-def}) or consider a constant $\chi=\chi_0$. In the former case, the carriers are strongly confined at the LAO/STO interface, while in the latter the carrier density shows a smooth and gradual decrease. More quantitatively, we can define the critical thickness $\xi$ as the distance at which the local carrier density is reduced by a factor of 1000 from the value just at the LAO/STO interface. Within the case of Eq.~(\ref{chi-def}) we find $\xi$=15~nm, while for $\chi=\chi_0=24000$, $\xi$=160 nm. We note that simulations with a constant $\chi=\chi_0=300$ (room-temperature case) yield $\xi \simeq$13 nm, in  agreement with Ref.~\onlinecite{siemons2007}.

Fig.~\ref{profiles}b shows that the strong carrier confinement obtained in the case of Eq. (\ref{chi-def}) is related to the presence of a very large electric field at the interface ($\sim$1 V.nm$^{-1}$). Note that although the electric field is indeed huge at the interface, the total potential rise does not exceed 600 meV (inset of Fig. \ref{profiles}b) for the fourth atomic planes at the vicinity of the LAO/STO interface. Remarkably, this potential rise amplitude matches well that found by recent first-principles calculations \cite{janicka09}. This strong electric field variation produces a giant decrease of $\chi$ close to the interface, as shown in Fig. \ref{profiles}c. When the variation of $\chi$ with the electric field is taken into account we find a value of $\xi$=15~nm that agrees very well with the metallic gas thickness of $\sim$12 nm measured at 8K by CTAFM (see Fig. \ref{ctafm}b and d). This definitely confirms that the field dependence of $\chi$ is an essential ingredient to describe the system at low temperature.

\begin{figure}[h!]
\centering
\includegraphics[width=\columnwidth]{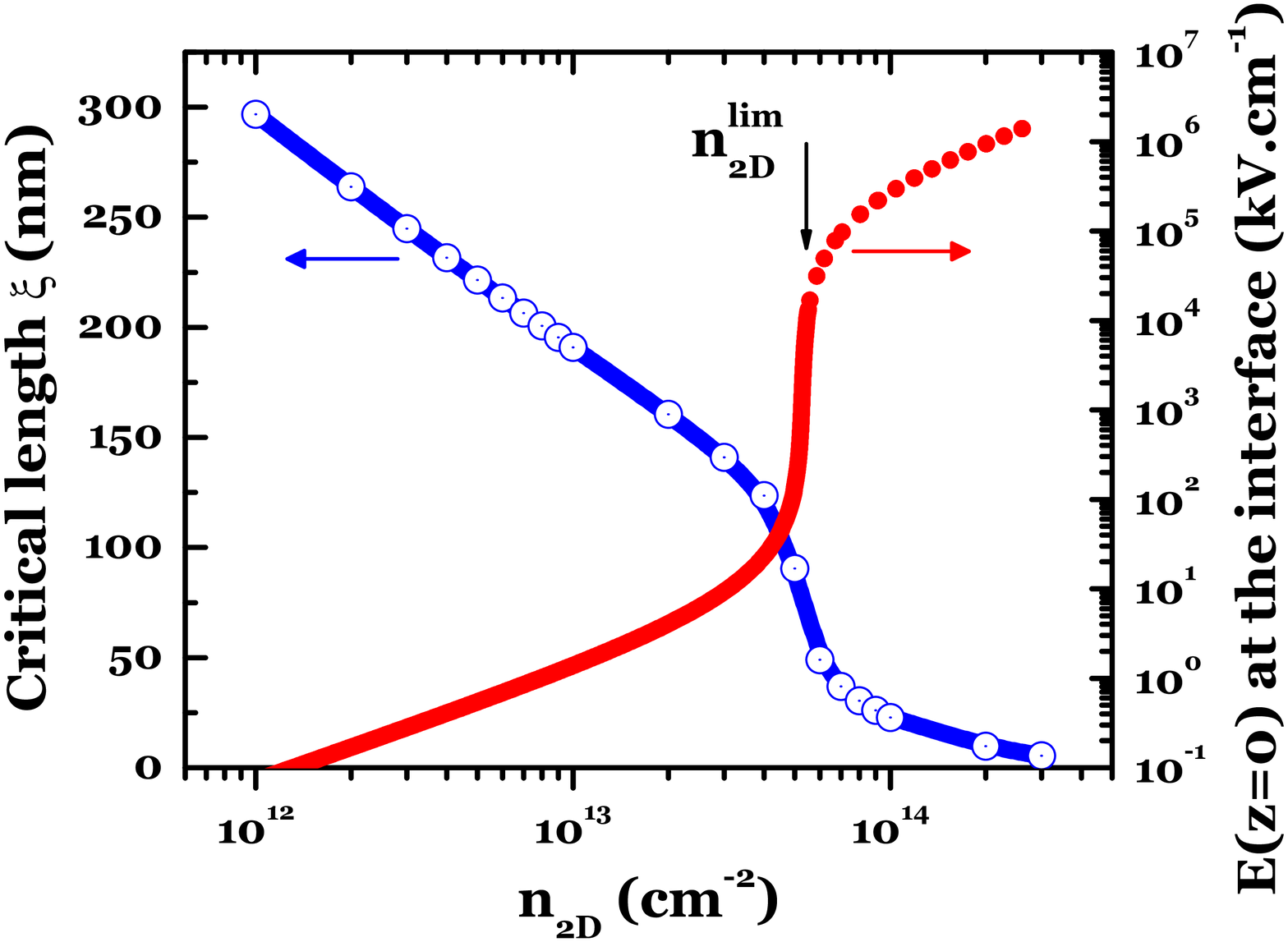}
\caption{(Color online) Dependence of the critical thickness $\xi$ and the maximum interfacial electric field with the sheet carrier density.}
\label{critical}
\end{figure}

To obtain a more general description of the LAO/STO system at low temperature, we have carried out similar simulations for a range of n$_{2D}$  including typical experimental values from the literature \cite{thiel2006,reyren2007,huijben2006,caviglia2008} and the theoretical value n$_{2D}$=$3.3\times 10^{14}$ cm$^{-2}$ \cite{ohtomo2004}. On Fig. \ref{critical} (right axis), we have plotted the electric field at the LAO/STO interface $\mathbb{E}^{z=0}$ as a function of $n^{2D}$. $\mathbb{E}^{z=0}$ rises sharply by more than three orders of magnitude beyond a certain sheet carrier density threshold $n_{2D}^{lim}=\mathcal{P}_{Sat}/e \approx 6\times 10^{13}$ cm$^{-2}$ and electric field $\mathbb{E}^{z=0}_{lim}=\mathcal{P}_{Sat}/(\epsilon_{0}\chi_{0}) \approx 4.7~kV$.cm$^{-1}$. This critical variation of $\mathbb{E}^{z=0}$ with n$_{2D}$ defines two regimes for $\xi$(n$_{2D}$) (Fig. \ref{critical}, left axis). When $\mathbb{E}^{z=0} < \mathbb{E}^{z=0}_{lim}$ (n$_{2D}$ $\lesssim 6\times 10^{13}$cm$^{-2}$), the carriers are spread over hundreds of nm due the efficient screening of the interface charges by the large permittivity. In contrast, for $\mathbb{E}^{z=0} > \mathbb{E}^{z=0}_{lim}$ (n$_{2D}$ $> 6\times 10^{13}$cm$^{-2}$), the giant increase of the electric field causes a strong reduction of the permittivity (as in Fig. \ref{profiles}c), which weakens the screening efficiency and confines the carriers over a few nm close to the interface.

The critical two-dimensional carrier density n$_{2D}=6\times 10^{13}$cm$^{-2}$ that delimits the confinement regime of the electron gas at the LAO/STO interface appears, at least, two times larger than the carrier density extracted from standard Hall measurements. This suggests that only a part of the carriers screening the electric field at the LAO/STO interface effectively take part to the electrical transport. This may indicate different mobilities within the \textit{t$_{2g}$} Ti bands or even some localization effects for a certain type of electrons~\cite{popovic2008}. Although this simple picture probably captures the main physics of the system, we note that (\textit{i}) the very high value of the electric field and (\textit{ii}) the low critical thickness obtained for large n$_{2D}$ values ($>10^{14}$~cm$^{-2}$) and extrapolated from the continuous limit would require a rigorous \emph{ab initio} treatment taking into account relaxation processes on atomic positions and atomic polarization~\cite{ishibashi2008,popovic2008,pentcheva2008,salluzzo2009}.

In summary, we have performed high and low-temperature conductive-tip AFM experiments in cross-section LAO/STO samples close to the interface. In both cases, we have found that the electron gas was confined over a typical thickness of $\sim$10 nm, in contrast with previous predictions of a much broader extension at low temperature~\cite{siemons2007}. We show theoretically that when the strong electric-field dependence of the permittivity in STO is taken into account the electron gas is indeed expected to be confined within a few nm both at low temperature and 300K. From the measured sheet carrier density and the electron gas thickness, we estimate the Fermi wavelength at low temperature to $\lambda_F \simeq$ 16 nm \cite{herranz2006b}, i.e. comparable to the gas thickness. The metallic gas is thus on the verge of two-dimensionality and it might be possible to induce a transition towards a pristine quantized state by adjusting the carrier density, e.g. with an electrostatic gate \cite{caviglia2008}.

\acknowledgments{We thank A. Caviglia, N. Reyren and J.-M. Triscone for useful discussions. We acknowledge financial support by the French ANR program (ANR Blanc "FEMMES" and "Oxitronics", ANR Pnano "Alicante"), the Triangle de la Physique and the Ile-de-France region (C-Nano program).}

\end{document}